\documentclass[a4paper,fleqn]{cas-sc}

\usepackage[numbers]{natbib}
\usepackage{physics}

\def\be{\begin{equation}}

\def\ee{\end{equation}}
\def\barr{\begin{array}}
\def\earr{\end{array}}

\def\l{\left}

\def\ed{\end{document}}

\def\cs{{\bf S}}

\def\tmp{\widetilde{m_p}}

\def\ed{\end{document}}
\begin{document}

\shorttitle{}    



\shortauthors{Shinde and Srivastava}
\title [mode = title]{Multipartite entanglement study in $fp$-shell nuclei}  

\author[1]{Rohit M. Shinde}
\cormark[1]
\ead{rohit_ms@ph.iitr.ac.in}
\cortext[1]{Corresponding author}
\affiliation[1]{organization={Department of Physics, Indian Institute of Technology Roorkee},
	city={Roorkee},
	citysep={}, 
	postcode={ 247667}, 
	country={India}}

\author[1]{Praveen C. Srivastava}[orcid=0000-0001-8719-1548]
\ead{praveen.srivastava@ph.iitr.ac.in}

\date{\today}
\begin{abstract}
In this study, we present multipartite entanglement results for the Ca and Ti isotopic chains obtained from nuclear shell-model wavefunctions. Our investigation focuses on the first excited $2^+_1$ states, which exhibit signatures of shell closures at $N=28$ and $N=32$. We calculate the 4- and 6-tangles, together with their corresponding network representations, providing a detailed picture of the underlying many-body correlations among nucleons. The primary objective of this work is to investigate the influence of shell closures on higher-order $n$-tangles, thereby gaining insight into the evolution of multipartite entanglement in medium-mass nuclei.

\end{abstract}
\maketitle

\section{Introduction}
The study of entanglement has been an active area of research in recent years, driven largely by its ability to quantify the quantum complexity observed in many-body systems \cite{Horodecki_2009, chuang, CJ_2023, Robin_2021}. Bipartite entanglement measures have been extensively studied in nuclear systems, revealing their evolution across shell closures and shape transitions \cite{Chen_2026, Xu_2026, Shinde_2026, Sarma_2026}. These studies have led to significant developments in many-body methods that leverage weakly entangled sectors to reduce basis dimensions \cite{CJ_2024, Perez_2026} and exploit large-scale entanglement to accelerate calculations, particularly when the entanglement structure is simple enough to be reproduced efficiently with classical methods \cite{Gottesman_2004, Robin_2025}. Multipartite entanglement, which characterizes correlations among more than two subsystems, has emerged as a powerful tool in this context \cite{Savage_2024, Savage_2023, Coffman_2000, Wong_2001}. Unlike bipartite measures, multipartite entanglement can reveal the collective correlations that govern the structure of complex quantum states, making it well-suited for probing the physics of strongly correlated nuclear systems.

The nuclear shell model provides a natural framework for such investigations \cite{Caurier_2005, Otsuka_2020, Subhrajit1, Subhrajit2}. Built upon a harmonic oscillator single-particle basis that can be expressed in Slater determinant form by means of second quantization, providing an easier partitioning of the many-body wavefunction. This structure makes the shell model a strong framework for studying entanglement between different single-particle states, as the basis gives clarity on how correlations are distributed across the system.

In this work, we present a study of multipartite entanglement for the $2^+_1$ states of the Ca and Ti isotopic chains in the $fp$-shell. For the shell-model calculations, we use the GXPF1A interaction \cite{Honma_2005}, which captures the shell closures at $N=28$ and $N=32$ extremely well. The primary goal of this study is to understand how higher-order quantum correlations evolve as shell closures are approached and how they are reflected in multipartite entanglement. Such studies can help us formulate strategies to optimize calculations near shell closures and to understand the behavior of entanglement in the excited states of medium-mass nuclei.

This article is organized as follows: in section \ref{multi}, we introduce the structure of the shell-model wavefunction and the theoretical framework of multipartite entanglement, followed by the results and conclusion in sections \ref{results} and \ref{con}, respectively.

\section{Theoretical framework of multipartite entanglement}
\label{multi}
The shell model Hamiltonian is represented in second-quantized form as 
\begin{equation}
H = \sum_i \epsilon_i \hat{a}_i^\dagger \hat{a}_i
+ \frac{1}{2} \sum_{i,j,k,l}
V_{ijkl}\,
\hat{a}_i^\dagger
\hat{a}_j^\dagger
\hat{a}_k
\hat{a}_l.
\label{eq:shell_hamiltonian}
\end{equation}
Here, the coefficients $\epsilon_i$ and $V_{ijkl}$ are the single-particle energies and two-body matrix elements, respectively. The $\hat{a}_i^\dagger$ and $\hat{a}_i$ represent the creation and annihilation operators corresponding to the single-particle states $\ket{i}\equiv\ket{n_i,l_i,j_i,m_i,\tau_i}$, where $n_i$ is the radial quantum number, $l_i$ and $j_i$ are the orbital and total angular momenta, and finally the $m_i$ and $\tau_i$ are the projections of total angular momenta and isospins. For our calculations, we consider a core of $^{40}$Ca along with valence protons and neutrons in the $fp$-model space comprising of $0f_{7/2}$, $1p_{3/2}$, $1p_{1/2}$, $0f_{5/2}$ harmonic-oscillator orbitals. All the calculations are performed using the GXPF1A interaction.

\begin{figure}
    \centering
    \includegraphics[width=0.65\textwidth]{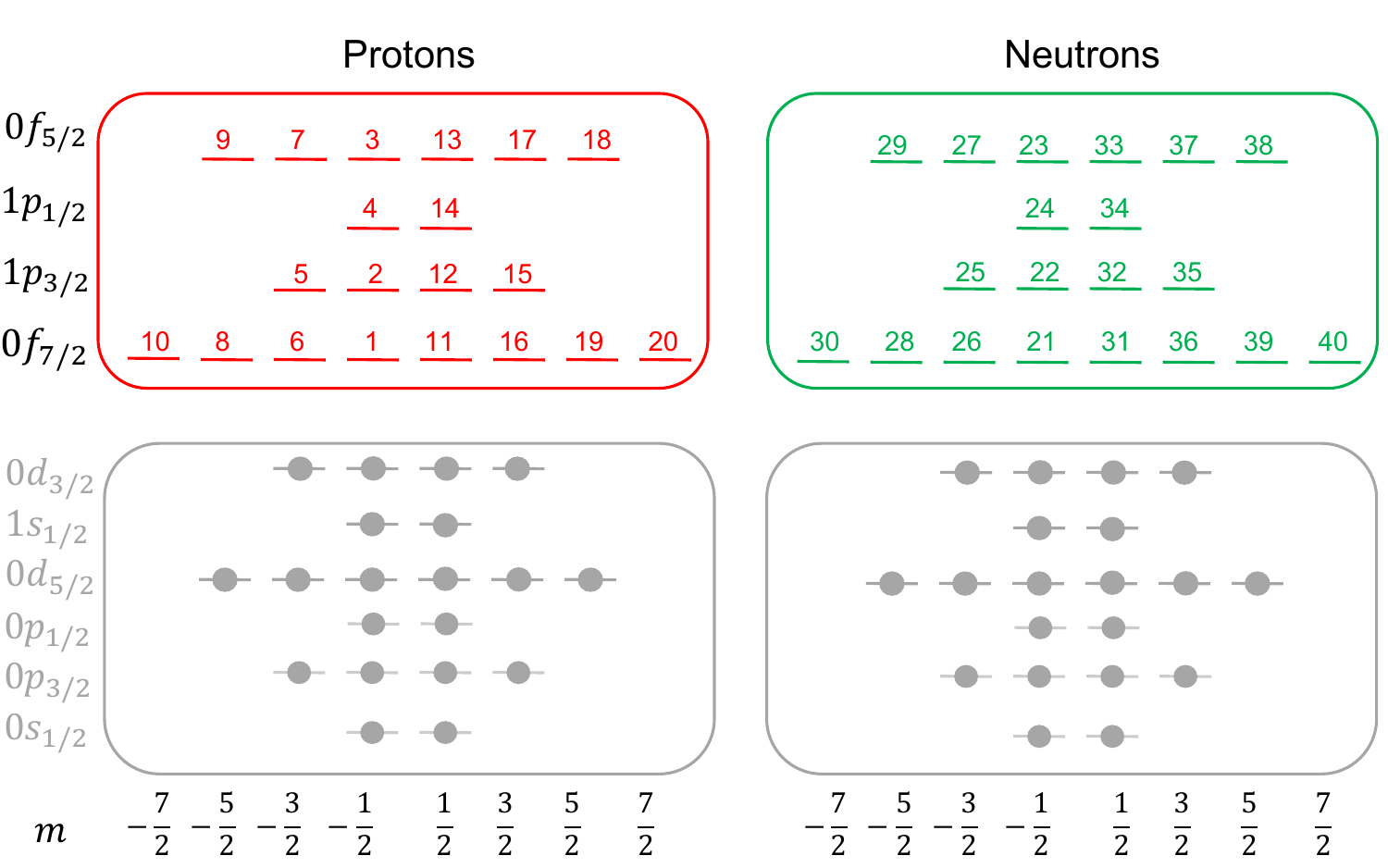}
    \caption{\small
    Representation of the single-particle states in the $fp$-shell model space. The fully filled $^{40}$Ca core is represented in gray.}
    \label{fig:pf_model}
\end{figure}

The many-body Schr\"odinger equation corresponding to the Hamiltonian in Eq.~(\ref{eq:shell_hamiltonian}) is solved by diagonalization in a basis of many-body Slater determinants. The resulting eigenstates are expressed as a linear combination of these basis states,
\begin{equation}
    |\psi\rangle = \sum_{\alpha} c_{\alpha} |\phi_{\alpha}\rangle ,
\end{equation}
where $c_{\alpha}$ are the expansion coefficients and $|\phi_{\alpha}\rangle$ denotes a many-body basis state in the occupation-number representation. In this formalism, each basis state is written as
\begin{equation}
|\phi_{\alpha}\rangle =
|n^{\alpha}_1 n^{\alpha}_2 \cdots n^{\alpha}_N\rangle,
\end{equation}
where $n_i^{\alpha}=0$ or $1$ specifies whether the $i^{\mathrm{th}}$ single-particle orbital is unoccupied or occupied, respectively. The calculations are performed in the $M$-scheme, where all basis states have a fixed total angular momentum projection. The calculations are conducted using the BIGSTICK code \cite{Johnson_2018, Johnson_2013}, the wave function is expressed using the single-particle states for $fp$-shell as shown in the Fig. \ref{fig:pf_model}.

Once the shell-model wavefunction is constructed in the occupation-number basis, each configuration
$|n^{\alpha}_1 n^{\alpha}_2 \cdots n^{\alpha}_N\rangle$
can be mapped onto an $N$-qubit using the Jordan-Wigner mapping, where occupancy of each single-particle orbital corresponds to the orientation of the mapped qubit. This correspondence allows the many-body correlations encoded in $|\psi\rangle$ to be investigated using tools from quantum information theory. Since shell-model wavefunctions naturally incorporate correlations among all valence nucleons within the chosen model space, multipartite entanglement measures provide a convenient way to characterize these collective quantum correlations. In the present work, we employ the $n$-tangle, which quantifies the entanglement shared among a selected set of $n$ qubits embedded within the larger $N$-qubit system \cite{Robin_2025, Wong_2001}.

The $n$-tangle $\tau^{(n)}$ is defined as
\begin{align}
    \tau^{(n)}_{(i_1 ... i_n)} &= |\langle \Psi | \hat{\sigma}_y^{(i_1)} \otimes ... \otimes \hat{\sigma}_y^{(i_n)}| \Psi^* \rangle|^2 \; ,
\label{eq:n-tangle}
\end{align}
where $\sigma_y^{(i_k)}$ is the ``spin flipping" Pauli matrix acting on qubit $i_k$. In this work, we restrict our analysis to the 4- and 6-tangles. Since the shell-model wavefunctions conserve both proton and neutron numbers, odd-order $n$-tangles vanish identically under this definition of multipartite entanglement and therefore do not contribute. To gain further insight into the origin of the multipartite correlations, we classify the summed $n$-tangle $(\overline{\tau}^{(n)})$ into proton-proton, neutron-neutron, and proton-neutron sectors given by

\begin{figure}
\centering
\includegraphics[width=0.47\textwidth]{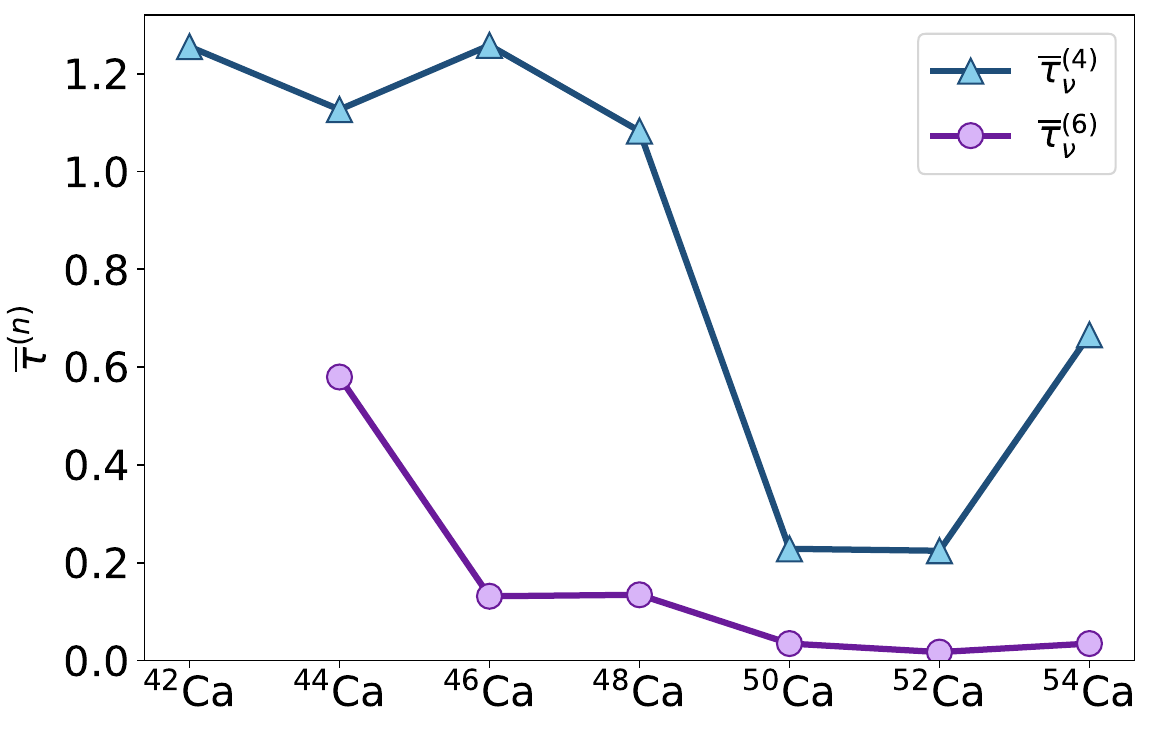}
\hfill
\includegraphics[width=0.48\textwidth]{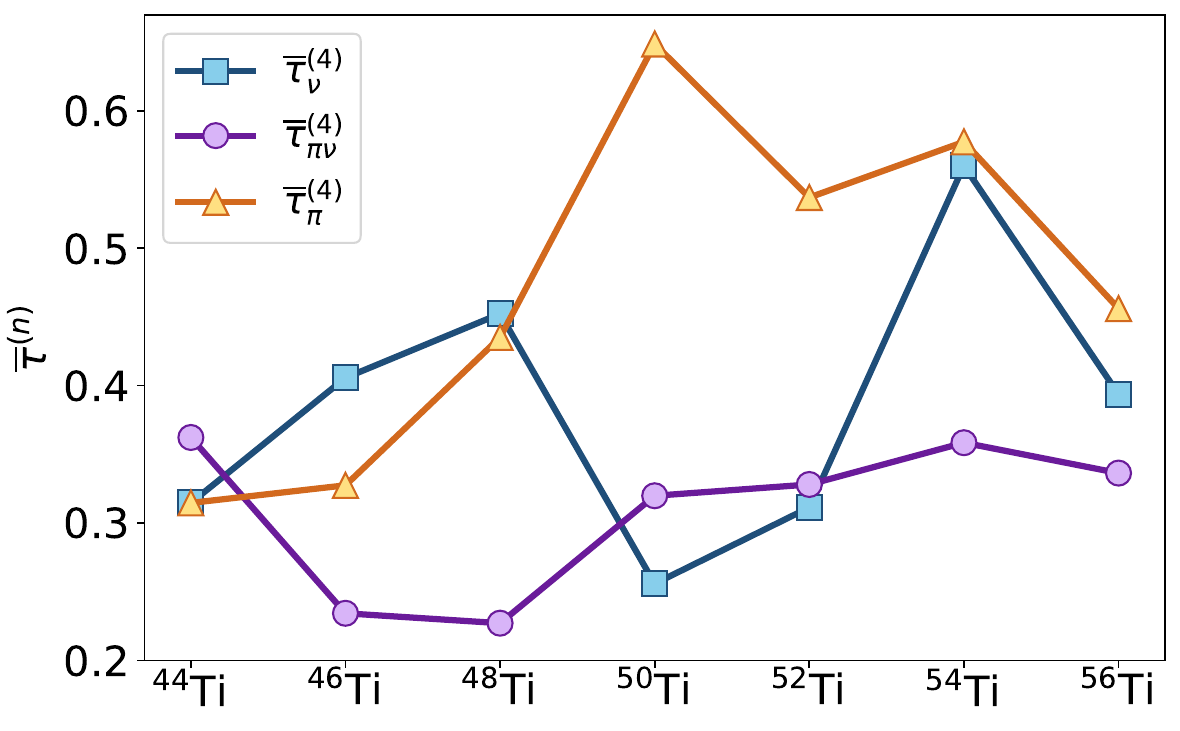}
\vspace{0.5cm}
\includegraphics[width=0.48\textwidth]{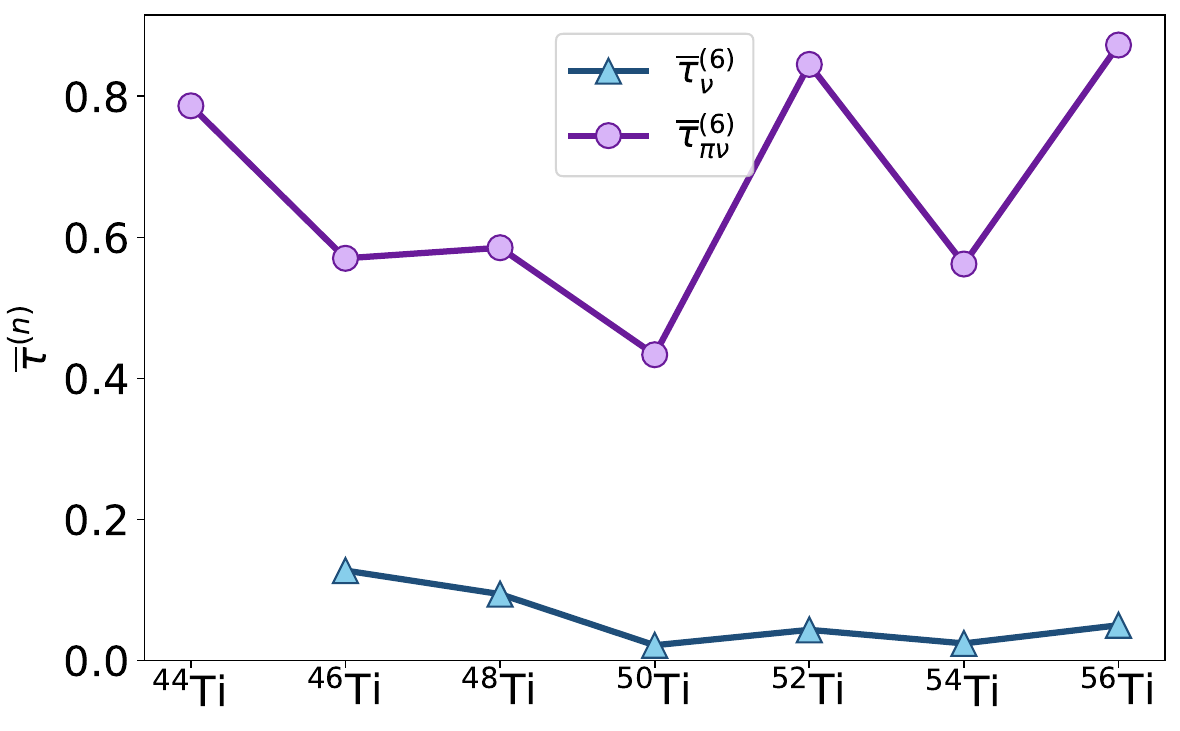}
\caption{\small Upper left panel: Values of the pure neutron summed 4- and 6-tangles for the Ca isotopic chain. Upper right panel: Values of the pure proton, pure neutron, and proton-neutron summed 4-tangles for the Ti isotopic chain. Lower panel: Values of the pure neutron and proton-neutron summed 6-tangles for the Ti isotopic chain.} 
\label{fig:combined_ntangle}
\end{figure}

\begin{equation}
\begin{aligned}
\overline{\tau}^{(n)}_{\pi}
&\equiv
\sum_{\substack{i_1,i_2,\ldots,i_n\\ \text{all protons}}}
\tau^{(n)}_{(i_1i_2\cdots i_n)},
\\[1.2em]
\overline{\tau}^{(n)}_{\nu}
&\equiv
\sum_{\substack{i_1,i_2,\ldots,i_n\\ \text{all neutrons}}}
\tau^{(n)}_{(i_1i_2\cdots i_n)},
\\[1.2em]
\overline{\tau}^{(n)}_{\pi\nu}
&\equiv
\sum_{\substack{i_1,i_2,\ldots,i_n\\ \text{mixed}}}
\tau^{(n)}_{(i_1i_2\cdots i_n)}.
\end{aligned}
\label{eq:summed_ntangles}
\end{equation}

While the summed $n$-tangles provide an overall measure of multipartite entanglement, they do not directly reveal how these correlations are distributed among the individual shell-model orbitals. To visualize this structure, we construct network representations in which each node corresponds to a single-particle orbital and the edges represent their joint participation in the multipartite correlations.

The edge connecting orbitals $i_1$ and $i_2$ is defined by summing the 4-tangle over all remaining pairs of orbitals,
\begin{align}
    e^{(4)}_{i_1 i_2} = \sum_{i_3 < i_4}   \tau^{(4)}_{(i_1, i_2, i_3, i_4)} \; ,
    \label{eq:edge_4tangle}
\end{align}
which provides a measure of the overall contribution of the orbital pair $(i_1,i_2)$ to the four-body entanglement of the many-body wavefunction. A similar representation can be extended to 6-tangles as well.

In the resulting network diagrams, the edge weight is represented by both the thickness and the color intensity of the connecting link, with larger values indicating stronger correlations. The node positions are obtained using a force-directed layout, causing strongly correlated orbitals to cluster together while weakly correlated orbitals are placed farther apart. These network representations provide an intuitive visualization of how multipartite correlations are distributed among the shell-model orbitals and how their structure evolves across the isotopic chains.


\begin{figure}
    \centering
    \includegraphics[width=0.98\textwidth]{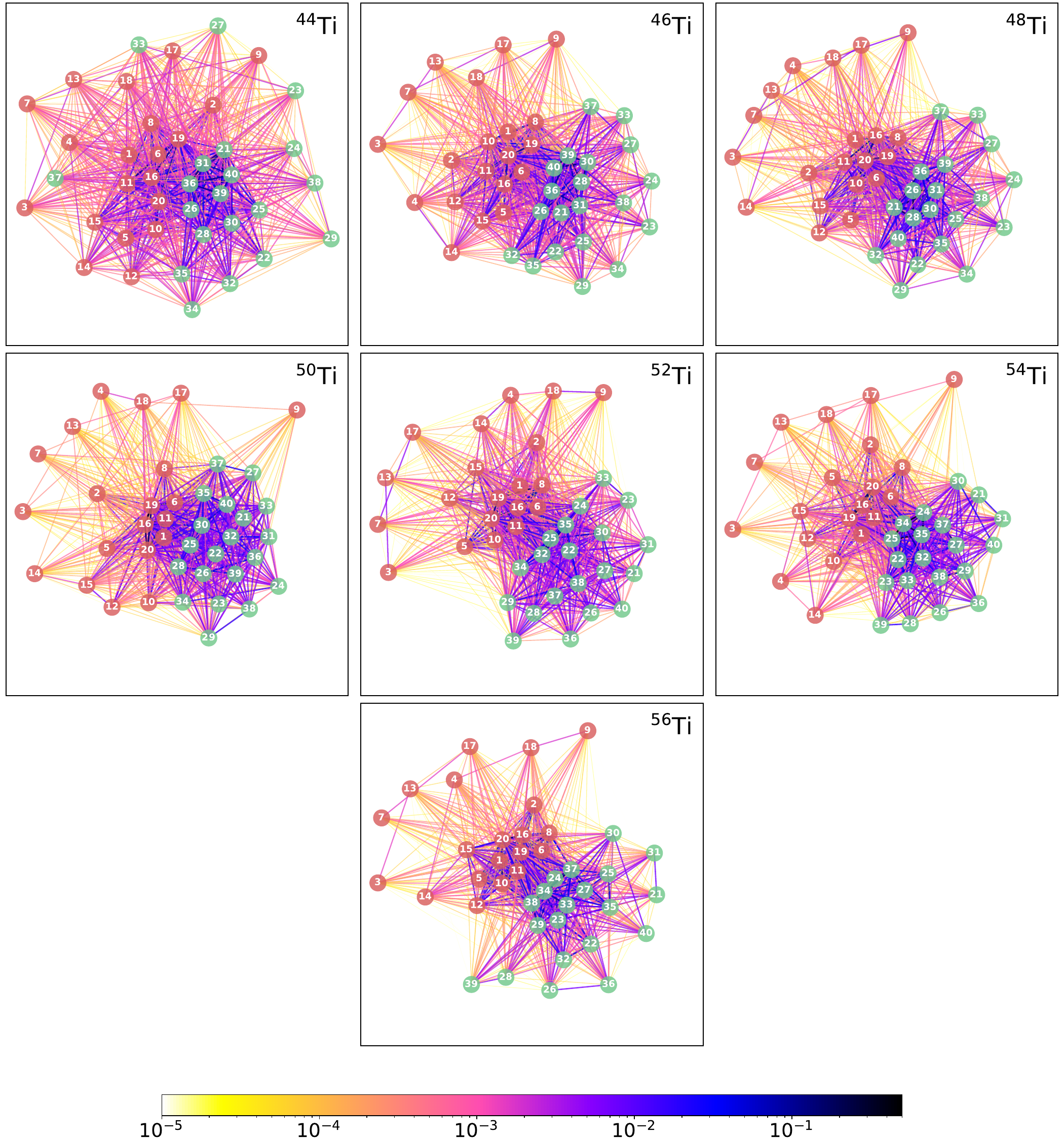}
    \caption{\small Network representation of the 4-tangles for the Ti isotopic chain. The nodes have been labeled according to the single-particle orbital in Fig.\ref{fig:pf_model}, where red nodes represent protons and green nodes represent neutrons. A logarithmic scale has been used to represent the contributions between two nodes because of large variations in entanglement. The NetworkX Python package is used to generate these plots \cite{Hagberg_2008}.
    }
    \label{fig:network_4tangle}
\end{figure}

\section{Results}
\label{results}
We start our discussion with the summed 4-tangles ($\overline{\tau}^{(4)}$) and summed 6-tangles ($\overline{\tau}^{(6)}$) for Ca and Ti isotopes as shown in Fig.~\ref{fig:combined_ntangle}. All calculations are performed for the $2^+_1$ states using the $J_z=J$ projection. For the Ca isotopes, there are no proton-proton or proton-neutron $n$-tangles, as there are no protons in the $pf$ model space. The $\overline{\tau}^{(4)}_\nu$ hover around 1.2 till $^{48}$Ca and then decreases significantly for $^{50}$Ca and $^{52}$Ca, then increases again for $^{54}$Ca. The $\overline{\tau}^{(6)}_\nu$ is highest for $^{44}$Ca and then decreases sharply for all the remaining neutron-rich nuclei. The Ti isotopes have all the contributions for the $\overline{\tau}^{(4)}$, where the $\overline{\tau}^{(4)}_\pi$ increases from $^{44}$Ti to $^{50}$Ti, and a small peak at $^{54}$Ti as well. For the $\overline{\tau}^{(4)}_{\pi\nu}$, $^{44}$Ti has a higher value, which suggests large correlations between protons and neutrons, correlations fall for the next two nuclei in the chain, and then slowly start rising. As for the $\overline{\tau}^{(4)}_{\nu}$, the neutron-neutron correlations rise till $^{48}$Ti, then it falls for $^{50}$Ti which coincides with the $N=28$ shell closure. A second maximum is observed at $N=32$, corresponding to the shell closure at $^{54}$Ti. The $\overline{\tau}^{(6)}_{\pi\nu}$ shows an interesting trend, $^{50}$Ti and $^{54}$Ti have a substantial dip in the proton-neutron correlations, which also corresponds to the shell closures. This suggests that higher-order $n$-tangles are sensitive to shell closures. The magnitude of $\overline{\tau}^{(6)}_{\nu}$ is lower than that of $\overline{\tau}^{(6)}_{\pi\nu}$, which also suggests that proton-neutron correlations are much stronger than neutron-neutron correlations.

In Fig.~\ref{fig:network_4tangle}, we have presented network plots for the 4-tangle corresponding to the Ti isotopic chain, where the strength of the edge corresponds to the contribution the two single-particle states (nodes) contribute to the 4-tangle for all possible configurations of the remaining single-particle states. For the $^{44}$Ti, the correlations are much stronger between the $f_{7/2}$ orbitals in the case of both protons and neutrons. For the $^{46}$Ti and $^{48}$Ti, we can observe that the proton $f_{5/2}$ and $p_{1/2}$ orbitals are not strongly entangled with the rest of the orbitals, but there is a strong correlation between proton $f_{7/2}$ and $p_{3/2}$. For the neutrons, the 4-tangles are strong even for the high-lying orbital as compared to proton orbitals. $^{50}$Ti shows a slightly stronger correlation between proton and neutron orbitals than the previous two nuclei, which can be seen in $\overline{\tau}^{(4)}_{\pi\nu}$ as well. As we move towards neutron-rich nuclei, the correlation between protons and fully filled neutron orbitals decreases, and they move away from the proton $f_{7/2}$ orbitals, which were closer at $^{44}$Ti.

\begin{figure}
    \centering
    \includegraphics[width=0.98\textwidth]{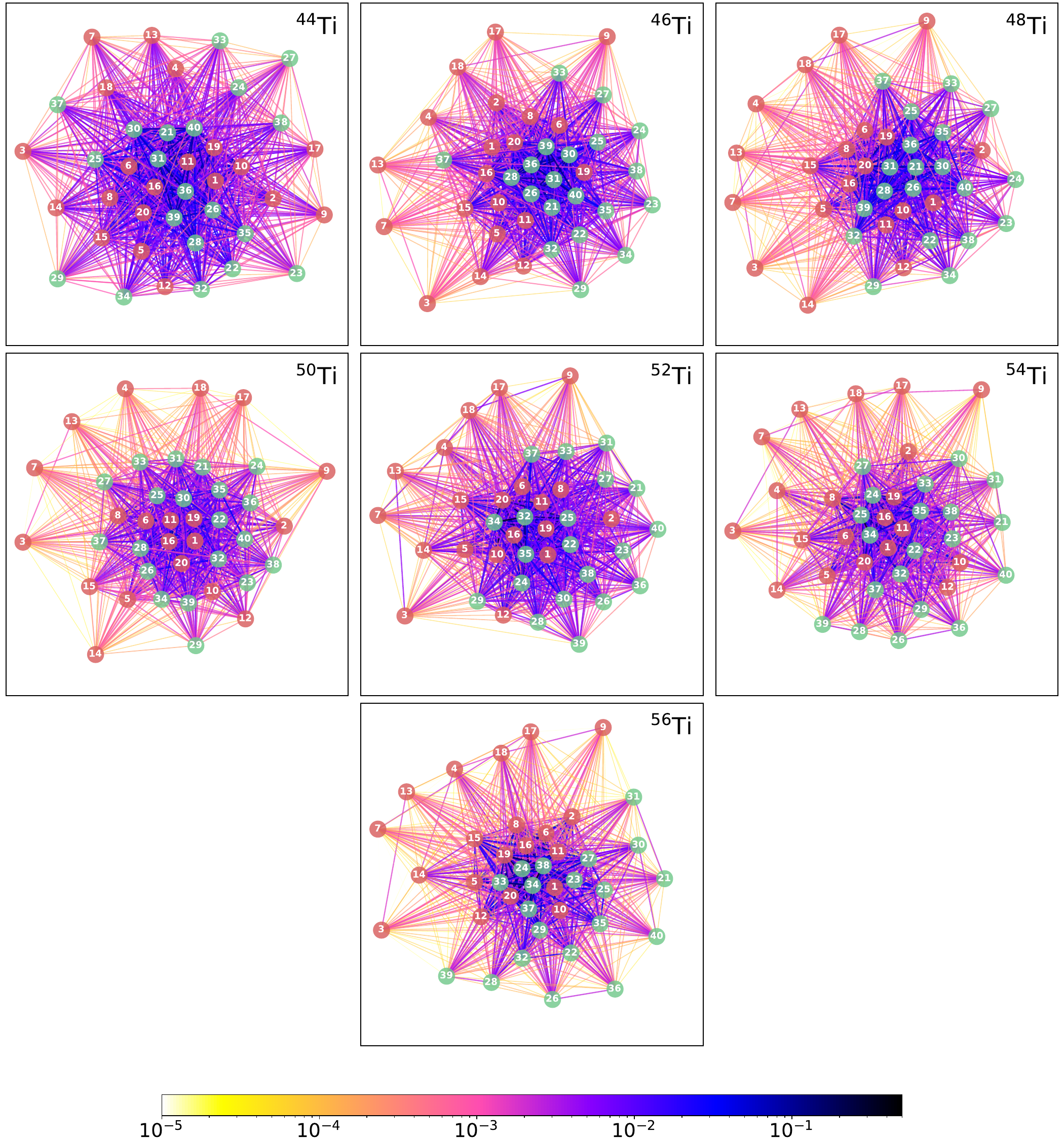}
    \caption{\small Network representation of the 6-tangles for the Ti isotopic chain. The nodes have been labeled according to the single-particle orbital in Fig.\ref{fig:pf_model}, where red nodes represent protons and green nodes represent neutrons. A logarithmic scale has been used to represent the contributions between two nodes because of large variations in entanglement.
    }
    \label{fig:network_6tangle}
\end{figure}

Fig.~\ref{fig:network_6tangle} represents the 6-tangle network plots for the Ti isotopic chain. The $N=Z$ nucleus shows that almost all the orbitals are strongly or moderately correlated with each other as compared to neutron-rich nuclei. For $^{46}$Ti and $^{48}$Ti, the correlations are considerably stronger between protons and neutrons as compared to 4-tangle correlations, where we can see that high-lying orbitals of protons are weakly entangled, which is also observed in 4-tangle correlations. For the $^{50}$Ti, the correlations are weak when compared to $^{52}$Ti, as reflected in the proton-neutron 6-tangles, $\overline{\tau}^{(6)}_{\pi\nu}$. The $^{56}$Ti again shows a strong correlation, but only between a few orbitals, unlike $^{46}$Ti and $^{48}$Ti, which can be due to the lower contribution from completely filled orbitals.

\section{Conclusion}
\label{con}
In this work, we have investigated multipartite entanglement in the $2^+_1$ states of the Ca and Ti isotopic chains within the $fp$-shell using the GXPF1A interaction. The 4- and 6-tangles reveal a clear evolution of many-body quantum correlations with increasing neutron number and exhibit distinct signatures near the $N=28$ and $N=32$ shell closures, particularly in the higher-order proton-neutron correlations. The network representation further illustrates how these correlations are distributed among the single-particle orbitals and how their structure changes as the shell closures are approached. These results demonstrate that multipartite entanglement provides a complementary perspective on nuclear structure, offering a sensitive probe of shell evolution and the underlying many-body correlations in medium-mass nuclei.

\section*{Acknowledgements}
We acknowledge financial support from MHRD (India) and research Grant No. ANRF/ARGM/2025/001130/TS from Anusandhan National Research Foundation [ANRF], India.  We would like to thank the National Supercomputing Mission (NSM) for providing computing resources of ‘PARAM Ganga’ at the Indian Institute of Technology Roorkee, implemented by C-DAC and supported by MeitY and DST, Government of India.




\end{document}